# *Reciprocal Quantum Electrodynamics with Bound States in the Continuum*

*Shoufeng Lan*[1,2,3*]

**Addresses:**

*[1] Department of Mechanical Engineering, Texas A&M University, College Station, TX 77843, USA*

*[2] Department of Materials Science and Engineering, Texas A&M University, College Station, TX 77843, USA*

*[3] Department of Electrical and Computer Engineering, Texas A&M University, College Station, TX 77843, USA*

Corresponding author information: shoufeng@tamu.edu

**Abstract:**

Quantum electrodynamics (QED) accurately describes all known forms of modern optics and photonics regarding interactions between photons and matter. While matter ranges widely from atoms, particles, to solids, photons are predominantly in a confined physical space, such as a pair of mirrors, for enhanced photon-matter interactions known as cavity QED. Since position and momentum are canonically conjugate variables governed by Heisenberg's uncertainty principle, a fundamental question arises − what if light confinement is in the not-so-intuitive momentum or reciprocal space? The realization of photonic bound states in the continuum (BICs) has made this exotic scenario possible. Here, we summarize the most recent advancements at this research frontier in optics and photonics, covering weak coupling, strong coupling, and nonlinear optics. We can designate such photon-matter interactions enabled by reciprocal light confinement through BICs with truly open systems as reciprocal QED, which holds great promise to comprehend and extend cavity QED for optics, photonics, and related fields.

## 1 Introduction

Historically, Dirac raised a divergence problem of QED [1], which states that the mass of an electron is infinite when using the non-relativistic Hamiltonian to describe a dynamic system of an atom amid "light quanta." Lamb then presented the famous "Lamb shift" with split energies that Dirac predicted to be degenerate and, importantly, a measured electron mass being finite [2]. These discrepancies led to the realization that the electron mass calculated by Dirac was not the electron mass measured by Lamb and that a renormalization procedure was necessary to link the two. Indeed, Tomonaga [3], Schwinger [4], and Feynman [5], who later shared the 1965 Nobel Prize, independently but equivalently realized this renormalization in a relativistic invariant theory [6]. Besides the theoretical framework, Haroche and Wineland found from an experimental standpoint that creating light confinement with a cavity played a central role in applying QED to manipulate the interactions between photons and matters down to a single photon or a single atom level [7, 8]. Such control of individual quantum systems could profoundly impact quantum information processing and many other fields, bringing on their Nobel Prize on cavity QED [9].

As cavity QED evolves, it has become one of the most instrumental techniques for studying and enhancing photon-matter interactions in coupled quantum systems such as condensed matter with cavity photons [10-15]. It is worth noting that sometimes classical interpretations also make great sense, for example, using two harmonic oscillators to model weak coupling, strong coupling, energy splitting, and anti-crossing [16]. These interpretations would not, however, overturn the fact that cavity QED underpins the phenomena. Under the umbrella of cavity QED, various optical structures, such as waveguides, Fabry-Pérot resonators, and photonic crystal (PhC) cavities, can provide the necessary confinement to generate cavity photons [17-19], thanks to advanced nanofabrication technologies driven by the contemporary electronic industry. Besides dielectric structures, metals, particularly noble metals (e.g., Au and Ag), in both plain and structured forms can also confine photons by coupling with electrons to form surface plasmon polaritons at the surface. Yet, all these confinements more or less rely on physical boundaries, for example, between a defect and the rest of PhC (**Fig. 1**, left, red line), at positions in real space. Under Heisenberg's uncertainty principle, in which position and momentum are canonically conjugate variables, the physical space is merely half of the available parameter space. And hence, an intuitive experimental quest is to look for light confinement in the other but largely unexplored half − the momentum or reciprocal space.

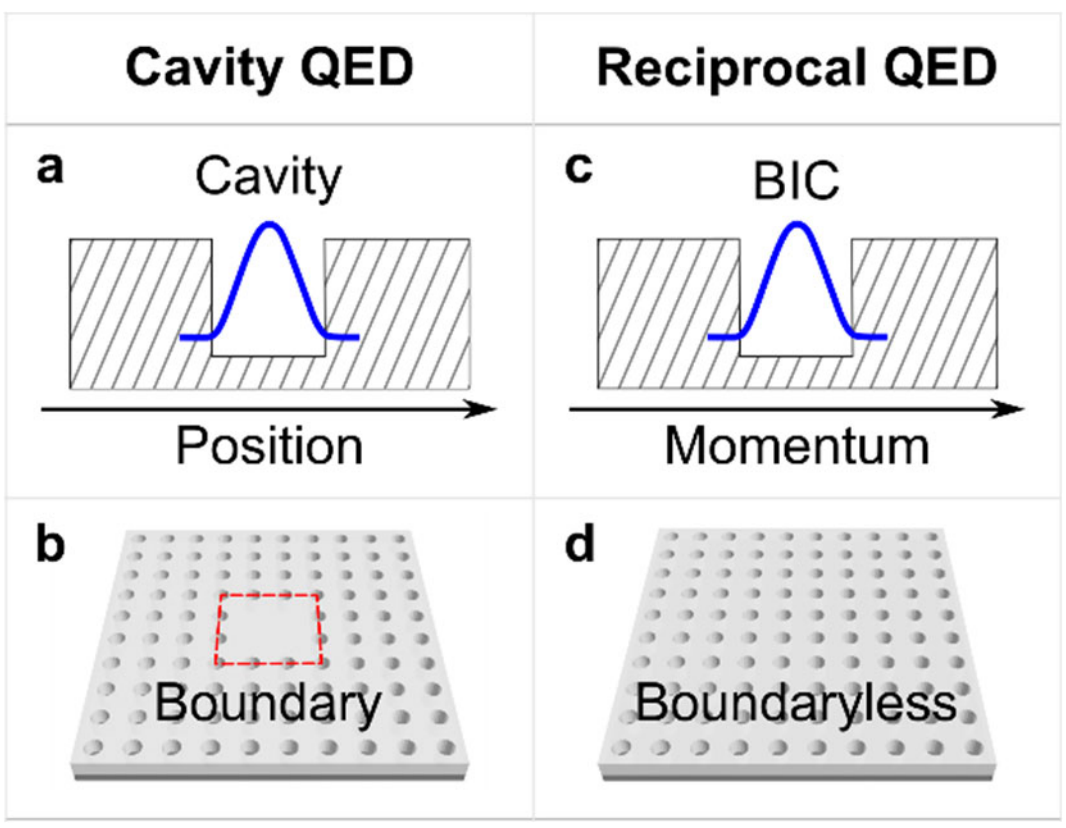

**Figure 1**: **Reciprocal and cavity QED both contain confined photons**. For cavity QED (left), the confinement (shaded area) is in the physical space with boundaries (dashed red line) surrounding a cavity. For the reciprocal QED (right), photon confinement based on a BIC is in the momentum space, with boundaryless or truly open optical structures.

Although photons in the momentum space, often known as Bloch photons [20], such as those in gratings and photonic crystals [21, 22], have dwelled for a long time, the realization of reciprocal confinement is not practical until the recent observation of BICs. The discovery of BICs, on the other hand, can date back to 1929 when von Neumann and Wigner found peculiar discrete solutions embedded in a continuous energy range for the Schrödinger equation in the context of quantum mechanics [23]. This mathematical concept became more physical with an interference theory among resonances introduced by Friedrich and Wintgen in 1985 [24]. Recently, researchers found that BICs manifest in zero linewidth or infinite lifetime can extent to wave phenomena governed by Maxwell's electromagnetism [25]. Following this, the experimental observations of photonic BICs in a one-dimensional waveguide array and PhCs finally came into reality [26, 27], and they stimulated a surge of renewed interest in fundamental research and practical applications [28-42]. Like the canonically conjugate pair of momentum and position, this emerging field named reciprocal QED (**Fig. 1**, right) deems to comprehend and extend the cavity QED by introducing BICs with light confinement in the momentum space.

## 2 Underline physical mechanisms of BICs

A BIC, also known as a quantum dark mode, is a nonradiative eigenstate situated within the energy range of a radiative continuum where it usually does not belong. One distinction we should note is that the BIC is not a Fano or any other resonances because they are leaky modes, but rather, the collapse of a resonance gives rise to a confined BIC. Phenomenologically, we can treat this collapsing behavior as a reciprocal potential well for photon confinement. It might not be intuitive, but an image of the potential well we can picture is that at least one characteristic of the system differs drastically from inside to outside the well, which forms a “wall.” For creating that “wall,” the characteristic used for a cavity is the effective refractive index (a local property), but for BICs, it can be, for example, symmetry (a global property). As a result, cavities confine photons in the real space localized within physical boundaries, while the BIC-based photon confinement is in the synthetic momentum space globalized to the entire device with truly open structures (**Fig. 1**). Another way to understand it is that confined photons for cavities are within the position range while those for BICs are within the momentum range of a potential well. Similar to a physical potential well for cavity QED, the reciprocal QED draws on a potential well in the momentum space achieved by BICs.

For photonic BICs, we can coarsely place them into symmetry-protected, accidental, and Friedrich-Wintgen categories [43]. Among them, the Friedrich-Wintgen BICs (**Fig. 2a**, ①) lean on an interference mechanism in which two or more radiative resonances destructively interfere, resulting in a nonradiative BIC. Because the resonances are part of the radiative continuum, the resulting BIC thus also situates within that energy range. Such a peculiar phenomenon is against a typical understanding of quantum mechanics, in which radiative and nonradiative solutions have separated energy levels. For a better visualization of the interference mechanism, **Fig. 2b** shows a full-wave simulation result of angle-resolved energy-momentum or *E*-*k* dispersion spectroscopy from a PhC with square lattices. The PhC supports two resonances (red and green dashed lines) with an avoided crossing after interfering with each other, leading to two new hybrid resonances (upper and lower). In the upper resonance, constructive interference results in a leaky mode, while destructive interference in the lower resonance collapses into a BIC (white star). The electric field distributions (**Fig. 2b**, right) illustrate unambiguously this interference process. The left and right parts of the upper resonance (UL and UR) share the polarity labeled as (+, −), so they are in phase

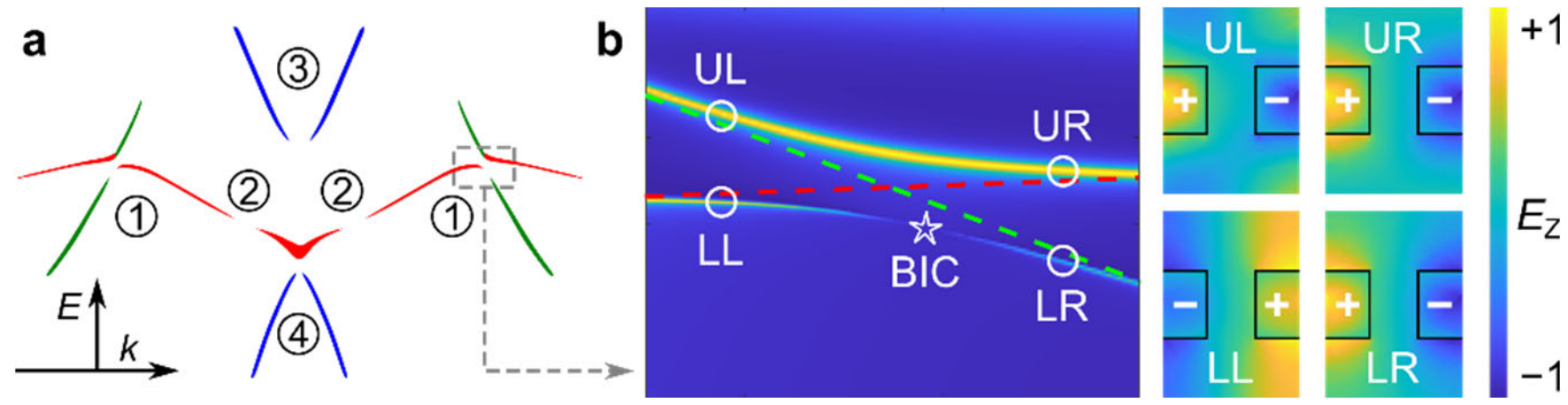

**Figure 2**: **Photonic BICs and the interference mechanism**. (a) ① is the Friedrich-Wintgen BIC,② is an accidental BIC, ③ and ④ are symmetry-protected BICs. (b) The coupling of two resonances (green and red dashed lines) leads to upper and lower hybrid modes. In the upper mode, the left (UL) and right (UR) parts corresponding to green and red resonances possess the same polarity (+, −). They are in phase and can constructively interfere, leading to a leaky mode. In the lower mode, the two parts have opposite polarities: (−, +) for the left (LL) and (+, −) for the right. Hence, they are out-of-phase and destructively interfere, resulting in a nonradiative BIC. (Adapted from Ref. [43])

and can constructively interfere. But for the lower resonance, the left part (LL) has a polarity of (−, +) while the right one (LR) is (+, −), so they are out-of-phase and can destructively interfere.

Another category of BICs is accidental (**Fig. 2a**, ②), achieved by fine-tuning the geometric parameters of nanostructures so that the transverse electric (TE with *p*-polarization) and magnetic (TM with *s*-polarization) modes accidentally cross the zero point simultaneously [27]. The zero point at specific $k$ values means no field couples to far fields or radiates out, thus forming a nonradiative BIC. Like Friedrich-Wintgen BICs, the accidental BICs are robust because small changes in the system parameters (e.g., periodicity) only move or tune the BIC to a different value of $E$ and $k$. Other than accidental BICs, symmetry-protected BICs can also obtain a zero-coupling to far fields because of a symmetry difference between radiative fields and the mode supported in the structure. Under time-reversal symmetry, a plane wave at an angle normal to the surface consists of a $C_2$ symmetry. The simplest way to obtain a symmetry difference is to have a lattice with higher-order symmetry, for example, $C_4$ in a square lattice. Because of the symmetry mismatch, it will prevent the mode in the structural lattices from coupling to far fields. That is why symmetry-protected BICs often occur at $k = 0$ or Γ-point, such as those in **Fig. 2a**, ③ and ④. Of course, the symmetry difference can also be accessible by purposely breaking or reducing the symmetry.

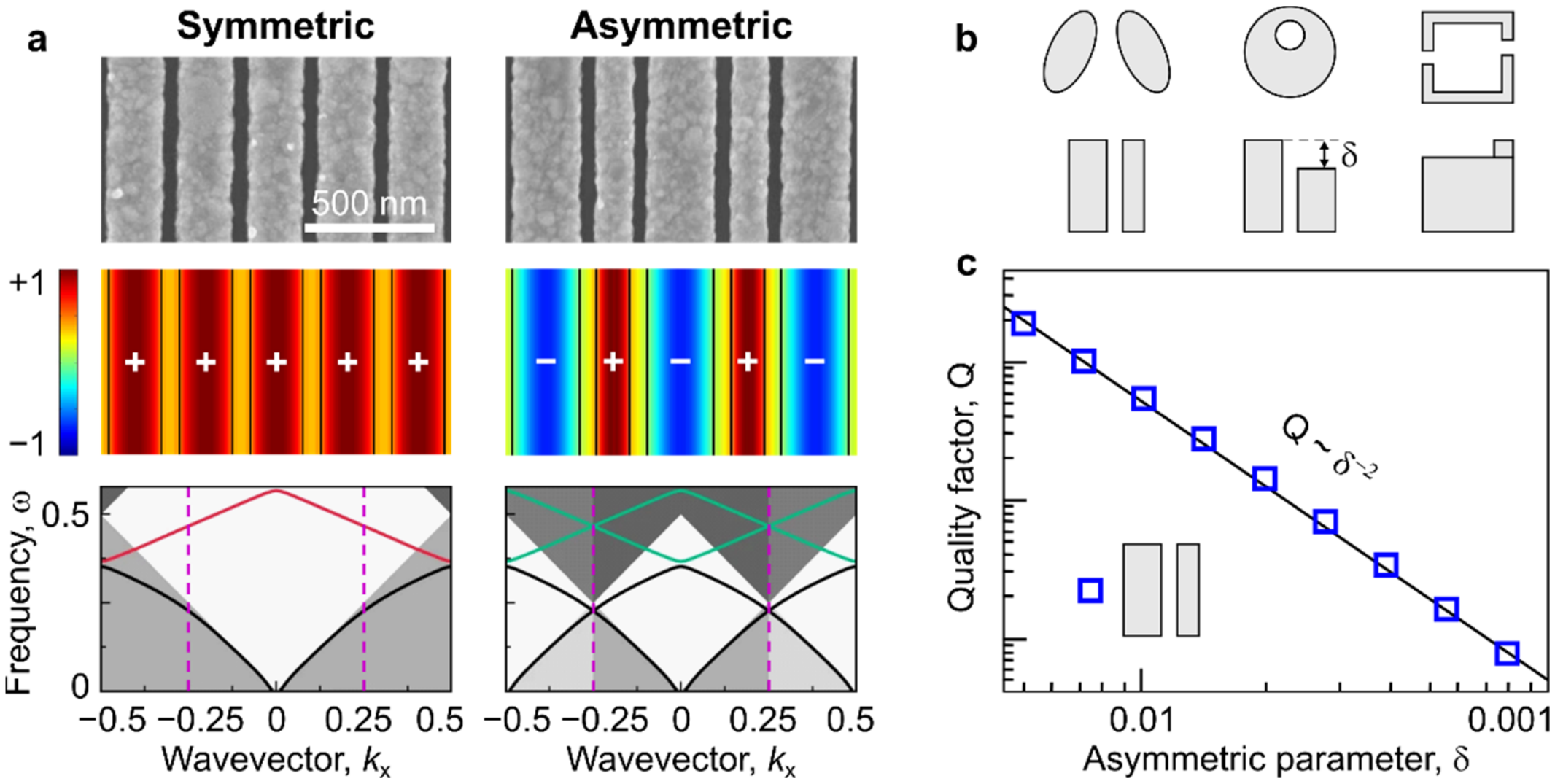

**Figure 3**: **Quasi-BICs achieved by purposely breaking the structrural symmetry**. (a) In 1D plasmonic crystals, the asymmetric structure has different widths for neighboring stripes. The symmetry breaking forces the electric fields to have opposite polarities (+ and −) and to destructively interfere with an intrinsically dark mode. Due to the band folding with a doubled periodicity and a halved wavevector (purple dashed line), the originally dark mode (lower right, black) is above the light line. The partially dark mode is a quasi-BIC that can couple to frar fields. (b) The symmetry breaking in 2D structures has more controllability by changing the shape, length, width, and area. (c) The quality factor, Q, of quasi-BICs, is proportional to $\delta^{-2}$, where δ is the asymmetric parameter. (Adapted from Refs. [44, 45])

BICs achieved by purposely reducing structural in-plane symmetries are quasi-BICs that host a finite lifetime or quality factor (Q) but are highly tunable. For example, **Fig. 3a** shows a one-dimensional plasmonic crystal that reduced the $C_{2v}$ symmetry by alternating the width of neighboring unit cells [44]. This symmetry-breaking forces the electric fields inside the adjacent unit cells to have opposite polarities (+ and −). As a result, they are out-of-phase and destructively interfere, forming a quantum dark mode. However, this dark mode is not fully nonradiative because it can couple to the far-field radiation through band-folding [46]. We can understand the band-folding technique by analyzing the first-order dark and bright optical modes. In the symmetric structure with identical unit cells, the first dark mode (black) is under the light line or not observable (**Fig. 3a**, bottom left). In asymmetric structures, the difference in neighboring unit cells doubles the periodicity, which, in turn, shrinks the momentum space in half. In other words, all modes in the photonic band structure will fold along the half of the maximum wavevector (purple dashed lines). After folding, the originally dark one (black) is in the white diamond or is observable, thus coupling to far-field radiations with a finite lifetime. As symmetry-breaking is general, the methodology can apply to two-dimensional (2D) structures (**Fig. 3b**) by changing the shape, location, width, length, or area [45]. It also applies to breaking higher-order symmetries, for example, $C_3$ and $C_4$, besides $C_2$ symmetry [47]. Most importantly, the asymmetric parameter of δ is tunable and determines the radiative $Q$ of the quasi-BICs with $Q = \delta^{-2}$ (**Fig. 3c**). For dissipative (e.g., plasmonic) systems, the total $Q$ will also consider the non-radiative or material Ohmic loss. With that said, the controllability of $Q$ equips quasi-BICs with another degree of freedom for manipulating photon-matter interactions.

## 3 BIC-enabled reciprocal QED for weak and strong coupling

Among photon-matter interactions, one of the best examples to illustrate QED is spontaneous emission, which is a quantum process that depends not only on the quantum emitter (matter) but also on the photonic environment with local density of states (photons) governed by Fermi's golden rule [48]. The photons conventionally refer to cavity photons in cavity QED, but in this review, they are BIC photons, otherwise specified, with confinement in the momentum space for the reciprocal QED. Under the semiclassical framework, only the description of the quantum emitter or matter part is in terms of quantum mechanics. For simplicity, we limit ourselves to a single-particle picture, in which the simplest model is a two-level system containing a ground state $|a\rangle$ and excited state $|b\rangle$ separated by a transition energy of $\hbar\omega_0$, where $\hbar$ is the reduced Plank constant. We can understand this transition process using a dipole with a dipole moment of $\boldsymbol{\mu}_{\mathrm{ba}} = \langle \mathrm{b}|\mathrm{q}\hat{\boldsymbol{r}}|\mathrm{a}\rangle$ and q being the elementary charge. We can then write down the Hamiltonian of the matter as $\mathcal{H}_{\mathrm{m}} = \hbar\omega_0\hat{L}^{\dagger}\hat{L}$, where $\hat{L} = |\mathrm{a}\rangle\langle\mathrm{b}|$ is a lowering operator. Subsequently, the interaction with photons in the environment, or BICs in this case, is a perturbation to the $\mathcal{H}_{\mathrm{m}}$.

On the other hand, the QED treats both matter and photons in the context of quantum mechanics, which makes it the most accurate theorem to describe all forms of photon-matter interactions in modern optics and photonics. The standard Hamiltonian of photons is $\mathcal{H}_{\mathrm{ph}} = \hbar\omega\hat{A}^{\dagger}\hat{A}$, where $\hat{A}$ is an annihilation operator and ω is the photon frequency, with quantized states being the vacuum state $|0\rangle$ and the Fock state $|n\rangle$ ($n \geq 1$). Now, both the dipole and electric field contribute to a photon-matter interaction with $\hat{\boldsymbol{\mu}} \cdot \hat{\boldsymbol{E}}$, where $\hat{\boldsymbol{\mu}} = \boldsymbol{\mu}_{\mathrm{ba}}(\hat{L}^\dagger + \hat{L})$ is the dipole moment operator, $\hat{\boldsymbol{E}} = E_{\mathrm{V}}(\hat{A}^\dagger + \hat{A})$ is the electric field operator, and $E_{\mathrm{V}}$ is the vacuum electric field. In the commonly made rotating wave approximation that ignores counter-rotating terms of $\hat{A}\hat{L}$ and $\hat{A}^\dagger\hat{L}^\dagger$, the Hamiltonian of the interaction is $\mathcal{H}_{\mathrm{int}} = \hbar \mathrm{g}(\hat{L}\hat{A}^\dagger + \hat{L}^\dagger\hat{A})$, where g is the coupling strength given by $\mathrm{g} = -\boldsymbol{\mu}_{\mathrm{ba}} E_{\mathrm{V}}/\hbar$. Overall, the Hamiltonian of the coupled system ($\mathcal{H}_{\mathrm{C}}$) contains three parts with $\mathcal{H}_{\mathrm{C}} = \mathcal{H}_{\mathrm{m}} + \mathcal{H}_{\mathrm{ph}} + \mathcal{H}_{\mathrm{int}}$, as detailed in the following Jaynes-Cummings (JC) equation [49, 50]

$$\mathcal{H}_{\mathrm{C}} = \hbar\omega_0 \hat{L}^\dagger\hat{L} + \hbar\omega\hat{A}^\dagger\hat{A} + \hbar\mathrm{g}(\hat{L}\hat{A}^\dagger + \hat{L}^\dagger\hat{A}) \quad (1)$$

Despite not fully decoupled from the other two parts, the last part of the JC equation or $\mathcal{H}_{\mathrm{int}}$ is the key to manipulating photon-matter interactions, which manifests in a single parameter, the coupling strength of g. First, because it involves both the dipole of $\boldsymbol{\mu}_{\mathrm{ba}}$ and the electric field of $E_{\mathrm{V}}$, it designates a general guideline for manipulating photon-matter interactions by working on either the matter or photon part. We can achieve the former by finding new emerging materials, such as two-dimensional and topological materials, and by designing exquisite experimental conditions, for example, cooling the system to cryogenic temperatures. For the latter or photon part, it provides greater flexibility with innovative degrees of control freedom by engineering photonic environments that surround quantum emitters [51], for instance, the creation of BICs in this review article. Second, depending on the strength of g, we can categorize photon-matter interactions in the weak, strong, and ultrastrong coupling regimes. In the ultrastrong coupling regime, g is comparable with $\omega_0$ (typically $2\mathrm{g} \geq 0.1\omega_0$), and the JC equation no longer holds because it needs to consider the counter-rotating terms [52]. It is a fascinating area that could lead to exotic phenomena, such as superradiance and the presence of virtual photons in the ground state [53-55]. By the time of this review article, to our knowledge, researchers have not achieved the ultrastrong coupling with BICs yet. Therefore, due to the limited space in this review, we will not discuss the details of the ultrastrong coupling regime, but rather, we will focus on the weak and strong coupling regimes shown in **Fig. 4**.

In the weak coupling regime (**Fig. 4**, **a-c**), g is smaller than the decay of the quantum emitter ($\gamma_{\mathrm{m}}$) and the decoherence of BIC photons ($\gamma_{\mathrm{ph}}$). Hence, the coupling with photons modifies the spontaneous emission from the quantum emitter, leading to a crossing effect in the spectrum. This crossing phenomenon with modified spontaneous emission best manifests in a Purcell factor [56],

$$F_{\mathrm{P}} = \frac{3}{4\pi^2}\left(\frac{\lambda_0}{n_{\mathrm{r}}}\right)^3 \frac{Q}{V} \tag{2}$$

where $\lambda_0$ is the wavelength of photons, $n_r$ is the refractive index of the environment, $Q$ is the quality factor of photon modes, and $V$ is the mode volume. The $Q$ is well-defined as $\omega/2\gamma_{\mathrm{ph}}$, while the $V$ extracted by the integration of normalized electric field distribution can be slightly different between non-dissipative (e.g., PhC with dielectric materials) and dissipative (e.g., plasmonic) modes [57]. Nevertheless, a general strategy to increase the $F_P$ factor is increasing $Q$ or decreasing $V$, ideal but challenging to achieve the two aspects simultaneously. For example, adding mirror-like structures, such as gratings and distributed Bragg reflectors (DBRs), or using materials with a high refractive index can shrink the $V$ for BICs and increase the g strength [58].

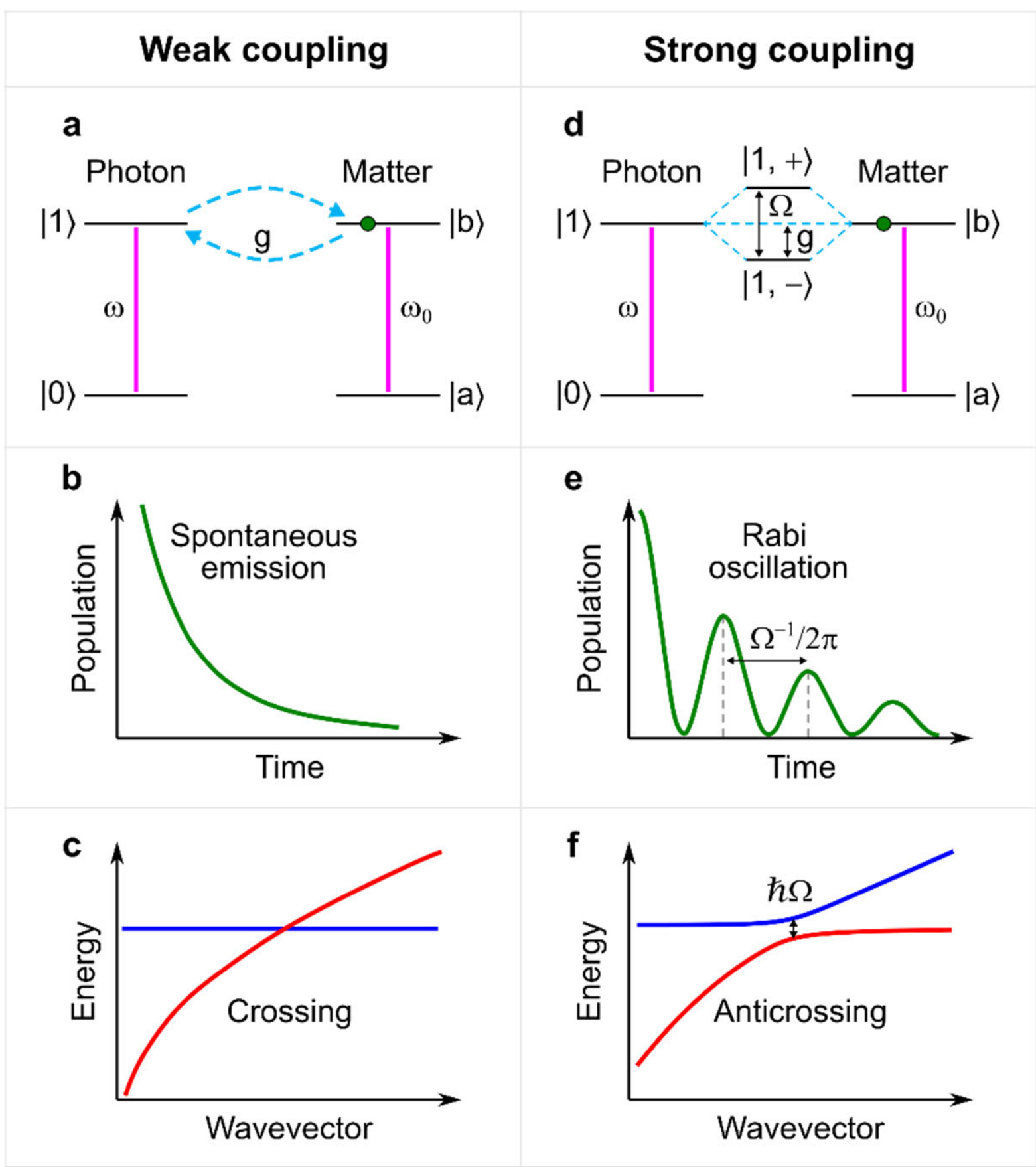

**Figure 4**: **The schematic illustration of weak and strong coupling governed by QED, in which both photon and matter manifest in the context of quantum mechanics**. For the weak coupling (**a-c**), the electron population (green dot) at the excited state |b⟩ gradually decays to the ground state |a⟩ through spontaneous emission with a spectral crossing effect. On the other hand, the strong coupling (**d-f**) results in two new dressed states |1, ±⟩ with the energy separation of Ω = 2g and an anticrossing effect provided by Rabi oscillation, where g is the coupling strength. |0⟩ and |1⟩ are the vacuum state and the first Fock state of the photon.

In the strong coupling regime (**Fig. 4**, **d-f**), g exceeds the $\gamma_m$ so that before the spontaneous emission fully decays the energy restores several times, resulting in a Rabi oscillation. Assuming in a resonant case $\omega = \omega_0$, the Rabi oscillation forces the coupled system out of the equilibrium and splits into two hybridized or half-photon, half-matter polaritonic states,

$$|n, \pm\rangle = \frac{|\mathrm{a}, n\rangle \pm |\mathrm{b}, n-1\rangle}{\sqrt{2}}, n \geq 1 \tag{3}$$

The energy levels of these polaritonic states are $E_n^{\pm}/\hbar = n\omega_0 \pm \mathrm{g}\sqrt{n}$, which shows a gap called Rabi splitting with $\Omega = (E^+ - E^-)/\hbar = 2\mathrm{g}\sqrt{n}$ between the upper and lower polaritonic states around the $n^{th}$ Fock state. The lowest order Fock state, also the most basic one, is with $n = 1$ shown in **Fig. 4**. As you can see, the above equation did not consider the loss for the quantum emitter and photons. But in reality, the dephasing and decoherence effects will always introduce losses ($\gamma_m$ and $\gamma_{ph}$), which modifies the energy accordingly. We can then roughly extract the modified energy levels (for $n = 1$) by diagonalizing the following non-Hermitian Hamiltonian,

$$\mathcal{H}_\mathrm{C} = \hbar \begin{pmatrix} \omega_0 - i\gamma_\mathrm{m} & \mathrm{g} \\ \mathrm{g} & \omega - i\gamma_\mathrm{ph} \end{pmatrix} \tag{4}$$

The resulting energy level of the two polaritonic states is

$$E_\pm/\hbar = \frac{\omega + \omega_0}{2} - \frac{i}{2}(\gamma_\mathrm{m} + \gamma_\mathrm{ph}) \pm \sqrt{\mathrm{g}^2 + \frac{1}{4}\left[\delta - i(\gamma_\mathrm{m} - \gamma_\mathrm{ph})\right]^2} \tag{5}$$

where $\delta = \omega - \omega_0$ is the detuning. Again, if we assume $\omega = \omega_0$ or in a resonant case, the Rabi splitting between the two states is $\Omega = \sqrt{4\mathrm{g}^2 - (\gamma_\mathrm{m} - \gamma_\mathrm{ph})^2}$, which has a real value only when $2\mathrm{g} > |\gamma_\mathrm{m} - \gamma_\mathrm{ph}|$. However, fulfilling this requirement does not guarantee an observable Rabi splitting with two peaks in photoluminescence or absorption spectra. For resolving the two peaks, the Rabi splitting has to exceed the linewidth of the polariton or $\Omega > \gamma_\mathrm{m} + \gamma_\mathrm{ph}$. Since $\gamma_m$ relates to the dephasing of the emission, lowering the temperature will limit the dephasing. Therefore, cooling the system to the cryogenic temperature mentioned before helps the observation of strong coupling. Similarly, increasing $Q$ and decreasing $V$ are also favorable for the strong coupling.

## 4 Nonlinear optical processes in BICs

Besides resonant states in which photons and matter are at the same energy level, non-resonant photon-matter interactions are also subject to the reciprocal QED. One prominent example is a nonlinear optical process, such as second and third-harmonic generation (SHG and THG). While matter stays at its ground state, photons change from an initial $|\mathrm{i}\rangle$ to the final state $|\mathrm{f}\rangle$ by going through a virtual state of annihilation $|\mathrm{v}\rangle$ and generation $|\mathrm{v'}\rangle$. Different from the semiclassical theorem, commonly being the default theory, by expanding the polarization into a Taylor series with respect to the applied electric field, reciprocal QED analyzes the transition between $|\mathrm{i}\rangle$ and $|\mathrm{f}\rangle$ states. As a result, the starting point and the best practice to study a nonlinear process governed by reciprocal QED is to define a BIC mode and representative material components involved and to identify their $|\mathrm{i}\rangle$ and $|\mathrm{f}\rangle$ states. To better illustrate such nonlinear processes, we can construct them with separable matter-photon product states, given by $|\mathrm{m}\rangle|\mathrm{ph}\rangle = |\mathrm{m};\mathrm{ph}\rangle = |E_{\mathrm{m}}^{\xi};\mathrm{N}(k,\eta)\rangle$, i.e., products of Hilbert states for the matter and Fock states for the photon [59]. Here, the number state of $\mathrm{N}(k,\eta)$ represents the electric field of photons, with N being the number of photons at the wavevector of $k$ and polarization index of $\eta$, where $k$ and $\eta$ together designate the BIC photon mode. For example, in SHG, the $|\mathrm{i}\rangle$ and $|\mathrm{f}\rangle$ states are $|E_{\mathrm{m}}^{\mathrm{a}};\mathrm{N}(k,\eta),0(k',\eta')\rangle$ and $|E_{\mathrm{m}}^{\mathrm{a}};(\mathrm{N}-2)(k,\eta),1(k',\eta')\rangle$, respectively, which denotes that the matter (subscript m) remaining in its ground state (superscript a) and two photons of one radiation mode converting into one photon of the other state.

Since the optical path in nanostructures for nonlinear photon-matter interactions is relatively short, focusing on the overall frequency conversion efficiency is crucial. To be consistent, we can continue using SHG, the most basic nonlinear process, as an example while not compromising the physics on the back. First, it is well-known that the generated new photon has a frequency of $\omega'$ double that of annihilated photons, inherently obeying the energy conservation, $\omega' = 2\omega$. Also, for the matter to keep at the ground state, it demands no change in the linear momentum of photons, enabling the wavevector-matching or phase-matching condition, $k' = 2k$. This condition is, nevertheless, largely relaxed in nanostructures due to their subwavelength thickness. Subsequently, the prevailing way for high conversion efficiency is to control the local density of states for $|\mathrm{i}\rangle$ and $|f\rangle$ states. Intuitive thinking is to design a BIC at the $|i\rangle$ state of the pumping light so that more free-space photons can convert into BIC photons, thus generating more nonlinear photons [60]. Likewise, creating additional BIC at the $|f\rangle$ state of nonlinear photons could also enhance the conversion efficiency [61]. The problem is that constructing two BICs at the $|i\rangle$ and $|f\rangle$ state by engineering the dispersion of materials is daunting. A promising approach for the double BICs is breaking the symmetry, for example, changing the shape of structures. By doing that, Doiron *et al.* achieved symmetry-guaranteed BIC pairs with tunable wavelengths [62]. Alternatively, one can develop a BIC at one state and a cavity at the other state. For instance, Wang *et al.* demonstrated the first BIC-cavity SHG with a remarkable conversion efficiency of $2.4\times10^{-2}$ $W^{-1}$ within a thickness less than 300 nm [63]. Interestingly, due to the BIC at the harmonic wavelength, they also observed a far-field vortex that carries an orbital angular moment.

The vortex or polarization singularity of BICs in the momentum space depicts its topological nature, which is intriguing due to the robustness against small perturbations in the annihilation, generation, and evolution processes [64]. Inspired by topological photonics [65-68], one possible way to annihilate a vortex, yet achieved in BICs, is to merge two with opposite vorticity or topological charges. Meanwhile, the most well-studied part is generating a BIC vortex with orbital angular moment by converting a photonic spin (circular polarization) through spin-orbit coupling. Specifically, to the reciprocal QED, the approach is through designing a BIC to resonate with the $|f\rangle$ state in various phenomena, such as diffraction [69, 70], photon emission [71], exciton-polariton [72, 73], and nonlinear optical generation [74]. Here, a nonlinear process is also the perfect platform to study the evolution of vortex by making a BIC resonate with the $|i\rangle$ state while detecting the vorticity of the $|f\rangle$ state. In view of that, Kang *et al.* numerically obtained an outstanding result that the evolved vorticity or quantum number of a BIC vortex in high harmonic generations follow a specific rule, which we can understand as angular phase matching [75].

Distinct from the regular phase matching governed by the conservation of linear momentum, angular phase matching is the conservation of angular momentum. Since the total angular mentum ($j$) consists of spin ($\sigma$) and orbital ($m_v$, also known as vorticity) counterparts, the angular phase matching manifests in a simple but powerful equation, $j_{\text{in}} = j_{\text{out}}$, where $j_{\text{in}}$ and $j_{\text{out}}$ are the input and output angular momentum (**Fig. 5a**). Though not named angular phase matching at that time, the first observation is in metasurface structures, where geometric phases in harmonic generations

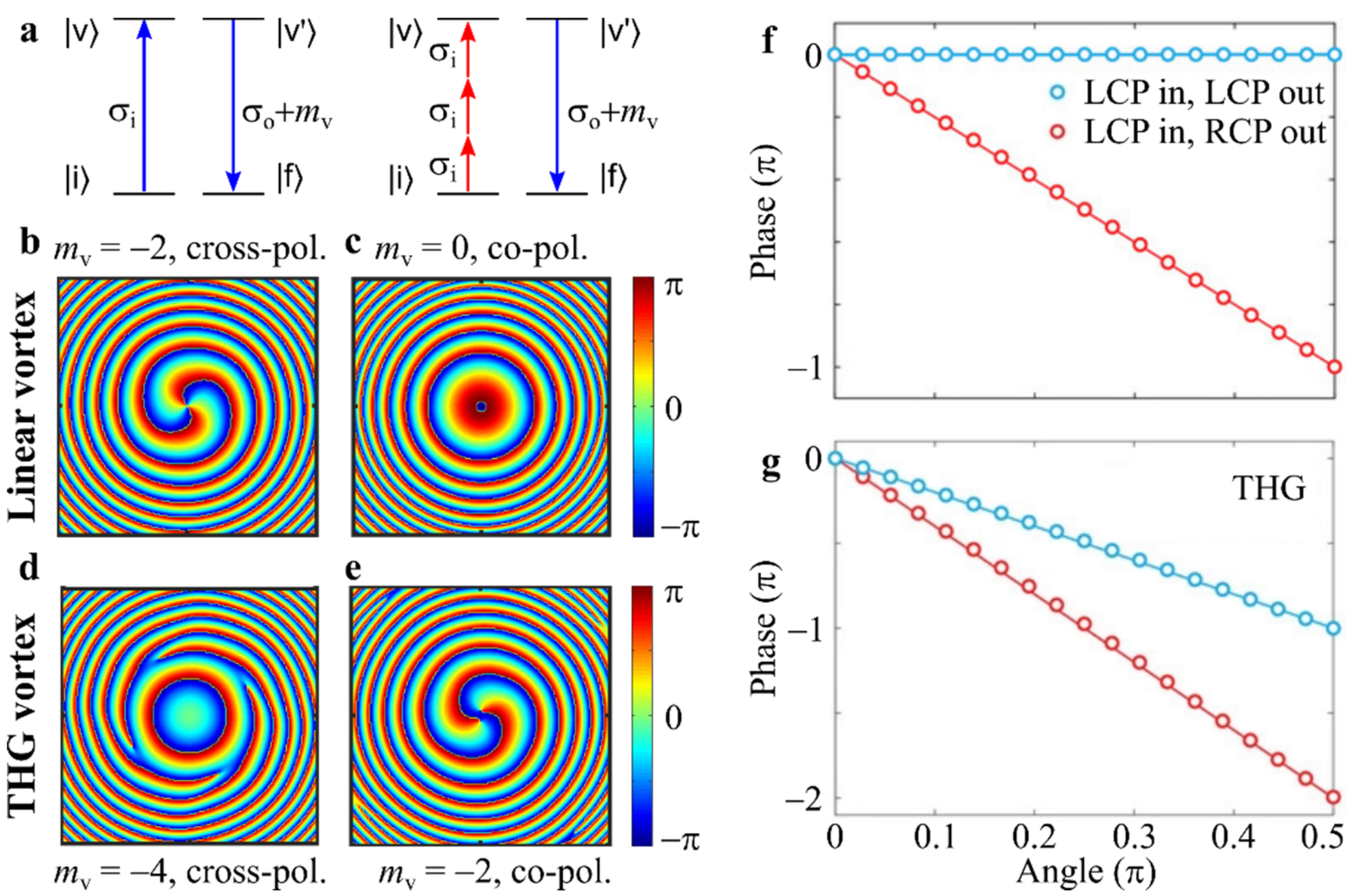

**Figure 5**: **Nonlinear optical vortex generation in BICs with angular phase matching**. (**a**) Governed by the angular momentum conservation, the input equals the output angular momentum, where $\sigma_i$, $\sigma_o$, and $m_v$ are input, output spin, and orbital angular momentum (vorticity). For left and right circular polarization (LCP and RCP), the σ is −1 and 1 accordingly. |i⟩, |v⟩, |v'⟩, and |f⟩ are initial, annihilation, generation, and final states. (**b**-**e**) The phase distribution in the momentum space for linear (**b** and **c**) and third-harmonic generation (THG) vortices using co- (LCP in, LCP out) and cross-polarizations (LCP in, RCP out). (**f** and **g**) The slope of curves with geometric phase (vertical axis) and polarization dipole angle (horizontal axis) designates the vortexity in linear and THG vortex generations. (Adapted from Ref. [75])

lead to continuous control of different $m_v$ values [76]. We can also understand the BIC vortex resulting from a geometric phase induced by polarization dipoles in a circle, and the $m_v$ designates how many turns a vortex resonates with the dipole at different angles. In other words, the $m_v$ is the slope of the curve for the generated geometric phase over the dipole angle (**Figs. 5f** and **5g**). This shared mechanism of the geometric phase enforces the BIC vortex to follow the same rule of angular phase matching as that for metasurfaces. For nonlinear processes, one thing to note is that the total angular momentum is a summation of all photons involved, for example, three input photons for THG. Also, it would be intriguing to verify whether the angular phase matching for BIC vortices follows the same selection rules on structural symmetry.

## 5 Discussion and perspectives

Despite the surging interest in BICs, the development of corresponding reciprocal QED is no more than at its infancy stage. Similar to any rising field, the foremost question is what differences it could bring, especially considering the cavity counterpart is well-studied. A quick answer is that the reciprocal QED offers new opportunities to bridge two communities with different or sometimes contradictory demands together. For the materials community, a simple form of photons is desirable for studying material physics and technology. For instance, the Fabry-Pérot microcavity consisting of two DBRs is the most widely adopted method for probing condensed matter with strong coupling. The problem is that the fabrication of DBRs that contain tens of nanolayers is by no means the easiest. The BICs can achieve similar functionality in the momentum space through planar periodic structures, which are fabrication-friendly. Thus, the reciprocal QED provides a new, simple format of photons while not compromising the desired functionality. For the photonics community, the interest lies in controlling and applying the generation, propagation, and detection of light that merges with simple materials as much as possible, for example, noble metals for surface plasmons. From this standpoint, the reciprocal QED with BICs opens new windows for integrating with other emerging materials, such as 2D materials, nanowires, and quantum dots, on the ground of open structures with photon confinement in the momentum space.

The reciprocal confinement also holds great promise for a new control of photon-matter interactions by revisiting the momentum degree of freedom. This parameter space contains both the linear and angular momenta, and the latter can further factorize into spin and orbital angular counterparts. Before reaching the full potential of momentum space, a critical process is to step back and understand how to probe and observe them. To that end, condensed matter physics set a great example that the linear momentum of electrons manifests in the electronic band structure and experimentally in angle-resolved photoemission spectroscopy (ARPES). Correspondingly, we can use photonic band structure and angle-resolved transmission, reflection, or absorption spectroscopy to interpret the linear momentum of photons. With this understanding, we own an enormous capacity to tailor linear momentum for various applications, for example, directional lasing at an oblique angle that is challenging for cavities. A similar methodology applies to angular momentum, for which 2D imaging in the reciprocal or *k*-space can reveal the orbital angular momentum of BICs with optical vortices or polarization singularities under the illumination of a circularly polarized light (the spin angular momentum). Although optical vortices and the related singular photonics have developed for a long time, there is still plenty of room for applications with reciprocal QED. One example is that an efficient nonlinear optical generation with reciprocal QED requires an overlooked angular phase matching governed by the conservation of spin and orbital angular momentum for the input and output light [75].

Nevertheless, we can list some unique properties of reciprocal QED and the roles they might play, and by no means is it a complete list because there must be many more we have not understood or even unforeseeable.

- Boundaryless structures with the potential to create large-area, single-mode, and scalable devices (e.g., surface-emitting lasers), compared to cavities with a limited size
- Open structures without the demanding need for being spatially deterministic, which could be essential for applications such as enhanced single-photon quantum emitters, compared to cavities with a strict requirement on location
- Multi-wavelength resonating in a single structure (e.g., crucial for low-threshold lasers to resonate at the pump and emission wavelengths), compared to a cavity with a single wavelength limited to material dispersion
- High directionality with an arbitrarily oblique angle for potential applications in LiDAR and laser-guided weapons, compared to cavities limited to the perpendicular direction
- Ultrahigh and tunable Q, for example, by controlling the degree of symmetric breaking (quasi-BICs), which could maximize enhancement with balanced radiative and nonradiative Q

The challenge, also the opportunity, for reciprocal QED to thrive is creating a systematic description of photon-matter interactions in a momentum language. To better sense what optical rules look like in the momentum language, we can take an inspiring demonstration of reciprocal lens for imaging as an example. Compared to conventional lenses that focus or bend light rays in the physical space, the reciprocal lens shifts rays through the extra phase created in the momentum space [77]. An open question is whether we have similar rules for BICs in reciprocal QED. In this context, the Purcell factor inversely proportional to the mode volume $V$ is a promising candidate because $1/V$ has the same unit with a three-dimensional wavevector that connects to the momentum by the Plank constant. Also, since a standing wave is an etalon benchmarking cavity QED, is there an equivalent analogous standing wave for reciprocal QED? This question might not be as wild as it appears when using Floquet states in semiconductors excited by optical vortices that carry orbital angular momentum [78]. Lastly, the reciprocal QED should stand on the shoulder of the cavity QED giant that laid a clear research roadmap. Can it be an alternative experimental platform to enable the measuring and manipulating of individual quantum systems as the Nobel Prize-winning cavity QED did? All in all, the materials and photonics communities must communicate closely to answer these fundamental open questions.

## Reference

[1] P. A. M. Dirac, "The quantum theory of the emission and absorption of radiation," *Proceedings of the Royal Society of London Series a-Containing Papers of a Mathematical and Physical Character,* vol. 114, no. 767, pp. 243-265, 1927.

[2] W. E. Lamb, and R. C. Retherford, "Fine structure of the hydrogen atom by a microwave method," *Physical Review,* vol. 72, no. 3, pp. 241-243, 1947.

[3] S. Tomonaga, "On a relativistically invariant formulation of the quantum theory of wave fields," *Progress of Theoretical Physics,* vol. 1, no. 1, pp. 27-42, 1946.

[4] J. Schwinger, "Quantum electrodynamics .1. A covariant formulation," *Physical Review,* vol. 74, no. 10, pp. 1439-1461, 1948.

[5] R. P. Feynman, "Space-time approach to quantum electrodynamics," *Physical Review,* vol. 76, no. 6, pp. 769-789, 1949.

[6] F. J. Dyson, "The radiation theories of Tomonaga, Schwinger, and Feynman," *Physical Review,* vol. 75, no. 3, pp. 486-502, 1949.

[7] M. Brune, F. Schmidt-Kaler, A. Maali *et al.*, "Quantum Rabi oscillation: A direct test of field quantization in a cavity," *Physical Review Letters,* vol. 76, no. 11, pp. 1800-1803, 1996.

[8] D. M. Meekhof, C. Monroe, B. E. King, W. M. Itano, and D. J. Wineland, "Generation of nonclassical motional states of a trapped atom," *Physical Review Letters,* vol. 76, no. 11, pp. 1796-1799, 1996.

[9] S. Haroche, "Nobel lecture: Controlling photons in a box and exploring the quantum to classical boundary," *Reviews of Modern Physics,* vol. 85, no. 3, pp. 1083-1102, 2013.

[10] J. M. Raimond, M. Brune, and S. Haroche, "Colloquium: Manipulating quantum entanglement with atoms and photons in a cavity," *Reviews of Modern Physics,* vol. 73, no. 3, pp. 565-582, 2001.

[11] H. Mabuchi, and A. C. Doherty, "Cavity quantum electrodynamics: Coherence in context," *Science,* vol. 298, no. 5597, pp. 1372-1377, 2002.

[12] R. Miller, T. E. Northup, K. M. Birnbaum, A. Boca, A. D. Boozer, and H. J. Kimble, "Trapped atoms in cavity qed: Coupling quantized light and matter," *Journal of Physics B-Atomic Molecular and Optical Physics,* vol. 38, no. 9, pp. S551-S565, 2005.

[13] H. Walther, B. T. H. Varcoe, B. G. Englert, and T. Becker, "Cavity quantum electrodynamics," *Reports on Progress in Physics,* vol. 69, no. 5, pp. 1325-1382, 2006.

[14] M. Pelton, "Modified spontaneous emission in nanophotonic structures," *Nature Photonics,* vol. 9, no. 7, pp. 427-435, 2015.

[15] F. Mivehvar, F. Piazza, T. Donner, and H. Ritsch, "Cavity QED with quantum gases: New paradigms in many-body physics," *Advances in Physics,* vol. 70, no. 1, pp. 1-153, 2021.

[16] L. Novotny, "Strong coupling, energy splitting, and level crossings: A classical perspective," *American Journal of Physics,* vol. 78, no. 11, pp. 1199-1202, 2010.

[17] H. Yokoyama, "Physics and device applications of optical microcavities," *Science,* vol. 256, no. 5053, pp. 66-70, 1992.

[18] K. J. Vahala, "Optical microcavities," *Nature,* vol. 424, no. 6950, pp. 839-846, 2003.

[19] R. Halir, P. J. Bock, P. Cheben *et al.*, "Waveguide sub-wavelength structures: A review of principles and applications," *Laser & Photonics Reviews,* vol. 9, no. 1, pp. 25-49, 2015.

[20] N. Rivera, and I. Kaminer, "Light-matter interactions with photonic quasiparticles," *Nature Reviews Physics,* vol. 2, no. 10, pp. 538-561, 2020.

[21] E. Yablonovitch, "Photonic band-gap structures," *Journal of the Optical Society of America B-Optical Physics,* vol. 10, no. 2, pp. 283-295, 1993.
[22] S. J. Smith, and E. M. Purcell, "Visible light from localized surface charges moving across a grating," *Physical Review,* vol. 92, no. 4, pp. 1069-1069, 1953.
[23] J. Von Neumann, and E. Wigner, "On some peculiar discrete eigenvalues," *Phys. Z.,* vol. 30, pp. 465-467, 1929.
[24] H. Friedrich, and D. Wintgen, "Interfering resonances and bound-states in the continuum," *Physical Review A,* vol. 32, no. 6, pp. 3231-3242, 1985.
[25] D. C. Marinica, A. G. Borisov, and S. V. Shabanov, "Bound states in the continuum in photonics," *Physical Review Letters,* vol. 100, no. 18, pp. 183902, 2008.
[26] Y. Plotnik, O. Peleg, F. Dreisow *et al.*, "Experimental observation of optical bound states in the continuum," *Physical Review Letters,* vol. 107, no. 18, pp. 183901, 2011.
[27] C. W. Hsu, B. Zhen, J. Lee *et al.*, "Observation of trapped light within the radiation continuum," *Nature,* vol. 499, no. 7457, pp. 188-191, 2013.
[28] C. W. Hsu, B. Zhen, A. D. Stone, J. D. Joannopoulos, and M. Soljacic, "Bound states in the continuum," *Nature Reviews Materials,* vol. 1, no. 9, pp. 16048, 2016.
[29] A. Kodigala, T. Lepetit, Q. Gu, B. Bahari, Y. Fainman, and B. Kante, "Lasing action from photonic bound states in continuum," *Nature,* vol. 541, no. 7636, pp. 196-199, 2017.
[30] L. Carletti, K. Koshelev, C. De Angelis, and Y. Kivshar, "Giant nonlinear response at the nanoscale driven by bound states in the continuum," *Physical Review Letters,* vol. 121, no. 3, pp. 033903, 2018.
[31] A. C. Overvig, S. Shrestha, and N. F. Yu, "Dimerized high contrast gratings," *Nanophotonics,* vol. 7, no. 6, pp. 1157-1168, 2018.
[32] A. Cerjan, C. W. Hsu, and M. C. Rechtsman, "Bound states in the continuum through environmental design," *Physical Review Letters,* vol. 123, no. 2, pp. 023902, 2019.
[33] K. Koshelev, S. Kruk, E. Melik-Gaykazyan *et al.*, "Subwavelength dielectric resonators for nonlinear nanophotonics," *Science,* vol. 367, no. 6475, pp. 288-292, 2020.
[34] S. I. Azzam, and A. V. Kildishev, "Photonic bound states in the continuum: From basics to applications," *Advanced Optical Materials,* vol. 9, no. 1, pp. 2001469, 2021.
[35] S. Joseph, S. Pandey, S. Sarkar, and J. Joseph, "Bound states in the continuum in resonant nanostructures: An overview of engineered materials for tailored applications," *Nanophotonics,* vol. 10, no. 17, pp. 4175-4207, 2021.
[36] J. H. Yang, Z. T. Huang, D. N. Maksimov *et al.*, "Low-threshold bound state in the continuum lasers in hybrid lattice resonance metasurfaces," *Laser & Photonics Reviews,* vol. 15, no. 10, pp. 2100118, 2021.
[37] Z. J. Yu, Y. Wang, B. L. Sun *et al.*, "Hybrid 2D-material photonics with bound states in the continuum," *Advanced Optical Materials,* vol. 7, no. 24, pp. 1901306, 2019.
[38] V. Kravtsov, E. Khestanova, F. A. Benimetskiy *et al.*, "Nonlinear polaritons in a monolayer semiconductor coupled to optical bound states in the continuum," *Light-Science & Applications,* vol. 9, no. 1, pp. 56, 2020.
[39] N. Bernhardt, K. Koshelev, S. J. U. White *et al.*, "Quasi-BIC resonant enhancement of second-harmonic generation in $ws_2$ monolayers," *Nano Letters,* vol. 20, no. 7, pp. 5309-5314, 2020.
[40] I. A. M. Al-Ani, K. As'Ham, L. J. Huang, A. E. Miroshnichenko, and H. T. Hattori, "Enhanced strong coupling of tmdc monolayers by bound state in the continuum," *Laser & Photonics Reviews,* vol. 15, no. 12, pp. 2100240, 2021.

[41] H. W. Yang, J. T. Pan, S. Zhang *et al.*, "Steering nonlinear twisted valley photons of monolayer ws$_2$ by vector beams," *Nano Letters,* vol. 21, no. 17, pp. 7261-7269, 2021.

[42] S. Cao, H. G. Dong, J. L. He, E. Forsberg, Y. Jin, and S. L. He, "Normal-incidence-excited strong coupling between excitons and symmetry-protected quasi-bound states in the continuum in silicon nitride-ws$_2$ heterostructures at room temperature," *Journal of Physical Chemistry Letters,* vol. 11, no. 12, pp. 4631-4638, 2020.

[43] X. Ma, K. Kudtarkar, Y. Chen *et al.*, "Coherent momentum control of forbidden excitons," *Nature Communications,* vol. 13, no. 1, pp. 6916, 2022.

[44] S. Lan, S. P. Rodrigues, M. Taghinejad, and W. Cai, "Dark plasmonic modes in diatomic gratings for plasmoelectronics," *Laser & Photonics Reviews,* vol. 11, no. 2, pp. 1600312, 2017.

[45] M. V. Rybin, K. L. Koshelev, Z. F. Sadrieva *et al.*, "High-Q supercavity modes in subwavelength dielectric resonators," *Physical Review Letters,* vol. 119, no. 24, 2017.

[46] W. H. Wang, Y. K. Srivastava, T. C. Tan, Z. M. Wang, and R. Singh, "Brillouin zone folding driven bound states in the continuum," *Nature Communications,* vol. 14, no. 1, pp. 2811, 2023.

[47] A. C. Overvig, S. C. Malek, M. J. Carter, S. Shrestha, and N. F. Yu, "Selection rules for quasibound states in the continuum," *Physical Review B,* vol. 102, no. 3, pp. 035434, 2020.

[48] M. O. Scully, and M. S. Zubairy, *Quantum Optics*, Cambridge: Cambridge University Press, 1997.

[49] E. T. Jaynes, and F. W. Cummings, "Comparison of quantum and semiclassical radiation theories with application to beam maser," *Proceedings of the Ieee,* vol. 51, no. 1, pp. 89-109, 1963.

[50] D. G. Baranov, M. Wersall, J. Cuadra, T. J. Antosiewicz, and T. Shegai, "Novel nanostructures and materials for strong light matter interactions," *Acs Photonics,* vol. 5, no. 1, pp. 24-42, 2018.

[51] X. Ma, N. Youngblood, X. Liu *et al.*, "Engineering photonic environments for two-dimensional materials," *Nanophotonics,* vol. 10, no. 3, pp. 1031-1058, 2021.

[52] A. A. Anappara, S. De Liberato, A. Tredicucci *et al.*, "Signatures of the ultrastrong light-matter coupling regime," *Physical Review B,* vol. 79, no. 20, 2009.

[53] P. Forn-Diaz, L. Lamata, E. Rico, J. Kono, and E. Solano, "Ultrastrong coupling regimes of light-matter interaction," *Reviews of Modern Physics,* vol. 91, no. 2, 2019.

[54] A. F. Kockum, A. Miranowicz, S. De Liberato, S. Savasta, and F. Nori, "Ultrastrong coupling between light and matter," *Nature Reviews Physics,* vol. 1, no. 1, pp. 19-40, 2019.

[55] J. J. Yu, J. Sloan, N. Rivera, and M. Soljačić, "Quantum electrodynamical metamaterials," *Physical Review A,* vol. 108, no. 3, pp. 033509, 2023.

[56] E. M. Purcell, "Spontaneous emission probabilities at radio frequencies," *Physical Review,* vol. 69, no. 11-1, pp. 681-681, 1946.

[57] C. Sauvan, J. P. Hugonin, I. S. Maksymov, and P. Lalanne, "Theory of the spontaneous optical emission of nanosize photonic and plasmon resonators," *Physical Review Letters,* vol. 110, no. 23, pp. 237401, 2013.

[58] E. Maggiolini, L. Polimeno, F. Todisco *et al.*, "Strongly enhanced light–matter coupling of monolayer ws$_2$ from a bound state in the continuum," *Nature Materials,* vol. 22, pp. 964-969, 2023.

[59] D. L. Andrews, D. S. Bradshaw, K. A. Forbes, and A. Salam, "Quantum electrodynamics in modern optics and photonics: Tutorial," *Journal of the Optical Society of America B-Optical Physics,* vol. 37, no. 4, pp. 1153-1172, 2020.

[60] C. Z. Fang, Q. Y. Yang, Q. C. Yuan *et al.*, "Efficient second-harmonic generation from silicon slotted nanocubes with bound states in the continuum," *Laser & Photonics Reviews*, 2022.

[61] T. Santiago-Cruz, S. D. Gennaro, O. Mitrofanov *et al.*, "Resonant metasurfaces for generating complex quantum states," *Science,* vol. 377, no. 6609, pp. 991-995, 2022.

[62] C. F. Doiron, I. Brener, and A. Cerjan, "Realizing symmetry-guaranteed pairs of bound states in the continuum in metasurfaces," *Nature Communications,* vol. 13, no. 1, 2022.

[63] J. Wang, M. Clementi, M. Minkov *et al.*, "Doubly resonant second-harmonic generation of a vortex beam from a bound state in the continuum," *Optica,* vol. 7, no. 9, pp. 1126-1132, 2020.

[64] B. Zhen, C. W. Hsu, L. Lu, A. D. Stone, and M. Soljacic, "Topological nature of optical bound states in the continuum," *Physical Review Letters,* vol. 113, no. 25, pp. 257401, 2014.

[65] M. Kim, Z. Jacob, and J. Rho, "Recent advances in 2D, 3D and higher-order topological photonics," *Light-Science & Applications,* vol. 9, no. 1, 2020.

[66] T. Ozawa, H. M. Price, A. Amo *et al.*, "Topological photonics," *Reviews of Modern Physics,* vol. 91, no. 1, 2019.

[67] Y. Ota, K. Takata, T. Ozawa *et al.*, "Active topological photonics," *Nanophotonics,* vol. 9, no. 3, pp. 547-567, 2020.

[68] D. Smirnova, D. Leykam, Y. D. Chong, and Y. Kivshar, "Nonlinear topological photonics," *Applied Physics Reviews,* vol. 7, no. 2, 2020.

[69] B. Wang, W. Z. Liu, M. X. Zhao *et al.*, "Generating optical vortex beams by momentum-space polarization vortices centred at bound states in the continuum," *Nature Photonics,* vol. 14, no. 10, pp. 623-628, 2020.

[70] Y. W. Zhang, A. Chen, W. Z. Liu *et al.*, "Observation of polarization vortices in momentum space," *Physical Review Letters,* vol. 120, no. 18, pp. 186103, 2018.

[71] S. Kim, B. H. Woo, S. C. An *et al.*, "Topological control of 2D perovskite emission in the strong coupling regime," *Nano Letters,* vol. 21, no. 23, pp. 10076-10085, 2021.

[72] V. Ardizzone, F. Riminucci, S. Zanotti *et al.*, "Polariton Bose-Einstein condensate from a bound state in the continuum," *Nature,* vol. 605, no. 7910, pp. 447-452, 2022.

[73] N. H. Dang, S. Zanotti, E. Drouard *et al.*, "Realization of polaritonic topological charge at room temperature using polariton bound states in the continuum from perovskite metasurface," *Advanced Optical Materials,* vol. 10, no. 6, 2022.

[74] M. Minkov, D. Gerace, and S. H. Fan, "Doubly resonant $\chi^{(2)}$ nonlinear photonic crystal cavity based on a bound state in the continuum," *Optica,* vol. 6, no. 8, pp. 1039-1045, 2019.

[75] L. Kang, Y. H. Wu, X. Z. Ma, S. F. Lan, and D. H. Werner, "High-harmonic optical vortex generation from photonic bound states in the continuum," *Advanced Optical Materials,* vol. 10, no. 1, pp. 2101497, 2022.

[76] G. X. Li, S. M. Chen, N. Pholchai *et al.*, "Continuous control of the nonlinearity phase for harmonic generations," *Nature Materials,* vol. 14, no. 6, pp. 607-612, 2015.

[77] W. Liu, J. Chen, T. Li *et al.*, "Imaging with an ultra-thin reciprocal lens," *Phys. Rev. X,* vol. 23, no. 13, 031039, 2023.

[78] H. Kim, H. Dehghani, I. Ahmadabadi, I. Martin, and M. Hafezi, "Floquet vortex states induced by light carrying an orbital angular momentum," *Physical Review B,* vol. 105, no. 8, pp. L081301, 2022.